\documentclass[aps,prb,reprint,groupedaddress,showpacs]{revtex4-1}

\usepackage{graphicx}
\usepackage{dcolumn}
\usepackage{bm}

\usepackage{amsmath}

\begin{document}
\title{Shot Noise Induced by  Electron-nuclear Spin-flip Scattering
\\ in a Nonequilibrium Quantum Wire}

\author{Kensaku Chida$^1$, Masayuki Hashisaka,$^{1, \dagger}$, Yoshiaki Yamauchi$^1$, Shuji Nakamura$^1$, Tomonori Arakawa$^1$, Tomoki Machida$^{2,3}$, Kensuke Kobayashi$^{1,}$}\email{kensuke@scl.kyoto-u.ac.jp}
\author{Teruo Ono$^1$}

\affiliation{$^1$Institute for Chemical Research, Kyoto University, Uji,
Kyoto 611-0011, Japan}

\affiliation{$^2$Institute of Industrial Science, University of Tokyo,
4-6-1 Komaba, Meguro-ku, Tokyo 153-8505, Japan}

\affiliation{$^3$Institute for Nano Quantum Information Electronics,
University of Tokyo, 4-6-1 Komaba, Meguro-ku, Tokyo 153-8505, Japan}

\affiliation{$^{\dagger}$Present address: Department of Physics, Tokyo Institute of Technology, Tokyo, Japan}

\begin{abstract}
We study the shot noise (nonequilibrium current fluctuation) associated with
dynamic nuclear polarization in a nonequilibrium quantum wire (QW) fabricated in
a two-dimensional electron gas. We observe that the spin-polarized conductance
quantization of the QW in the integer quantum Hall regime collapses when the QW
is voltage biased to be driven to nonequilibrium. By measuring the shot noise,
we prove that the spin polarization of electrons in the QW is reduced to $\sim
0.7$ instead of unity as a result of electron-nuclear spin-flip scattering.  The
result is supported by Knight shift measurements of the QW using resistively
detected NMR.
\end{abstract}

\date{\today}
\pacs{76.70.Fz, 73.43.Fj, 73.63.Nm, 76.60.Cq, 73.50.Td}


\maketitle

Nuclear spins in solids are well isolated from their environment, and their
coherence times can be much longer than that of electron spins. This fact makes
them a promising candidate as qubits~\cite{Kane1998nat} and
memories~\cite{Taylor2003PRL,McCameyScience2010} in solid-state devices.
Dynamic nuclear polarization (DNP) is the hyperfine-mediated transfer of spin
polarization from electrons to nuclei, and can be used to initialize the nuclear
spins before a computation~\cite{McCameyPRL2009}. Quantum wires (QWs) defined in
a two-dimensional electron gas (2DEG) in the quantum Hall (QH) regime could be
used for quantum information processing since coherent control of the nuclear
spin state has been demonstrated~\cite{Corcoles2009PRB,Yusa2005nat}. DNP in
these systems is triggered by voltage biasing the edge
states~\cite{WaldPRL1994,MachidaAPL2002} but the detailed mechanism of DNP in a
small confined region such as a QW is not fully understood.

Because DNP is expected to be compensated by the spin degree of freedom of
electrons, the resultant electron depolarization further provides essential
information on the nature of the electron-nuclear interaction.  The Knight shift
measurement has been applied to address DNP by resistively detected NMR (RDNMR)~\cite{Stern2004PRB, Kumada2007PRL, Masubuchi2006APL}, although, to the best of our knowledge, the direct observation of the electron-nuclear spin-flip process remains to be demonstrated. Shot noise (nonequilibrium current fluctuation) could enable us to further address DNP because it is generated by the partition process of electrons at scatterers.  Actually, shot noise enabled us to quantitatively discuss \ electron-spin-dependent transmission
probabilities~\cite{DiCarloPRL2006,Nakamura2009PRB}, which is impossible through
conventional conductance measurements alone~\cite{Blanter2000PR}.  To the best
of our knowledge, however, DNP has not been investigated by using shot noise.

The purpose of this Rapid Communication is to demonstrate that shot noise can serve to investigate the electron-nuclear interaction in nonequilibrium nanostructures. We prove that the shot noise is induced by the electron-nuclear
spin-flip scattering process in a nonequilibrium QW in the integer QH regime. By
combining the conductance and noise measurements, we deduce finite electron spin
depolarization as a result of DNP, which is confirmed to be consistent with the
result of the Knight shift measurement via RDNMR.

\begin{figure}[bp]
\center \includegraphics[width=.82\linewidth]{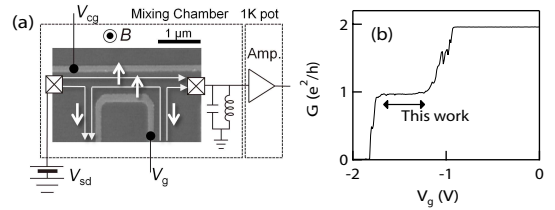} \caption{(a)
Schematic diagram of the measurement setup with a SEM image of a QW fabricated
on GaAs/AlGaAs 2DEG.  (b) Equilibrium QW conductance as a function of $V_g$.  At
the conductance plateau at $e^2/h$, which we focus on in this paper, only the
outer edge channel transmits electrons whereas the inner one is reflected.  The
two channels are spin polarized, as schematically shown in (a) by $\uparrow$ and
$\downarrow$.  }
\end{figure}

Figure~1(a) shows a schematic measurement setup with the scanning electron
microscopy (SEM) image of a QW fabricated on a AlGaAs/GaAs heterostructure 2DEG
[with an electron density of $2.3 \times 10^{11}$~cm$^{-2}$ and mobility of $1.1
\times 10^6$~cm$^2$/Vs]. The 1-$\mu$m-long QW is defined by two metallic gate
electrodes by applying gate voltages $V_g$ and $V_{cg}$~\cite{comment2}. We used
a resonant circuit with a homemade cryogenic amplifier for the noise measurement
and a one-turn coil around the device for RDNMR~\cite{Stern2004PRB}. The
source-drain bias voltage $V_{sd}$ is applied to the QW to simultaneously
measure the dc current, the differential conductance, and the noise in the
two-terminal geometry.  The noise spectral density centered around 2.5 MHz is
obtained as reported before~\cite{HashisakaRSI2009}.  By applying a magnetic
field ($B$) perpendicularly to the 2DEG, we tune the bulk 2DEG at the QH state
with a filling factor $\nu \approx 2$.  The base electron temperature ($T$) in
the QW, estimated by the thermal noise measurement, was 80~mK. In this
condition, the electron Zeeman energy $\Delta_Z = \vert g^* \mu_B B\vert \sim
120$~$\mu$eV at 5~T is much larger than the thermal energy ($k_B T\sim
7$~$\mu$eV at 80~mK), where $g^*$ is the \(g\) factor of bulk GaAs, $\mu_B$ is
the Bohr magneton, and $k_B$ is the Boltzmann constant. 

Figure~1(b) shows the equilibrium conductance of the QW as a function of $V_g$.
The conductance plateau at $2e^2/h$ for $V_g \gtrsim -0.9$~V reflects the $\nu =2$ QH state in the bulk 2DEG.  Because the edge states are fully spin polarized,
the QW conductance shows another plateau at $e^2/h$ between $V_g = -1.25$ and
$-1.75$~V, where only the outer edge channel passes through the QW with the
inner one perfectly reflected, as schematically presented in Fig.~1(a).  Below
we focus on this single-edge regime.

\begin{figure}[tbp]
\center \includegraphics[width=.83\linewidth]{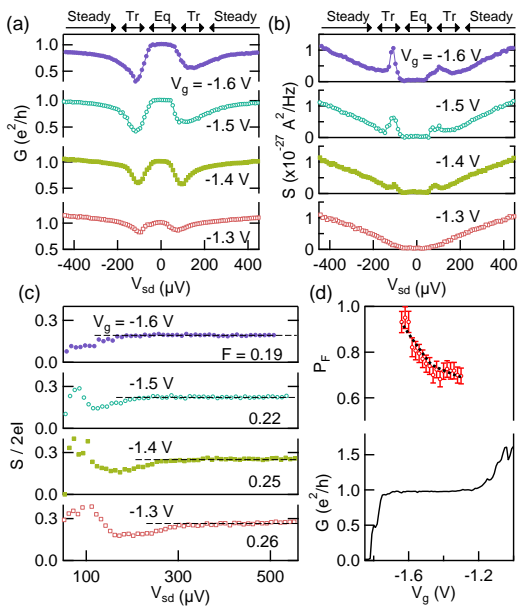} \caption{(Color
online) (a) Differential conductance $G$ at 4.5~T as a function of $V_{sd}$
obtained for $V_g =-1.6$, $-1.5$, $-1.4$, and $-1.3$~V.  (b) Corresponding
excess noise $S$. In the top part of (a) and (b), the (quasi-)equilibrium,
transient, and steady regimes are indicated by the arrows labeled ``Eq,''
``Tr,'' and ``Steady,'' respectively.  (c) Corresponding Fano factor
$S/2eI$. The straight dashed curves indicate the Fano factor in the steady
regime. (d) The spin polarization in the steady regime obtained by the noise
measurement $P_{F}$ (upper panel) and the equilibrium conductance as a function
of $V_g$ (lower panel). The dashed curve in the upper panel is a guide to the
eye.}
\end{figure}

Figure~2(a) shows the differential conductance $G$ at 4.5~T as a function of
$V_{sd}$ obtained for $V_g =-1.6$, $-1.5$, $-1.4$, and $-1.3$~V.  The behavior
of $G$ can be divided into three regions depending on the value of $|V_{sd}|$,
namely, the ``(quasi-)equilibrium,'' the ``transient,'' and the ``steady''
regimes.  In the small-$|V_{sd}|$ region [(quasi-)equilibrium], $G$ remains
constant at $e^2/h$.  When $|V_{sd}|$ is increased beyond a certain threshold
voltage, typically $50$--$100$~$\mu$V depending on $V_g$, $G$ starts to deviate
from the quantized value and shows a clear dip structure (transient regime).  In
this regime, $G$ shows hysteresis behavior depending on the sweep direction of
$V_{sd}$ and exhibits temporal variation over minutes, as we discuss later in
Fig.~3. [Note that the curves in Fig.~2(a) were obtained by sweeping $V_{sd}$ in
the backward direction.]  The threshold value of $|V_{sd}|$ for entering this
transient regime falls at the same order as $\Delta_Z$ and almost linearly
increases with $B$. Therefore, the observed collapse of the conductance
quantization is related to the electron scattering between Zeeman-split Landau
levels. Consistent with this fact is that when the filling factor is set to one
for both the bulk and the QW, the collapse of the conductance quantization is
observed to occur at much larger $|V_{sd}|$.  The upper bound of $|V_{sd}|$ for
the transient regime is determined by the broadening width of Landau levels.  By
further increasing $|V_{sd}|$ (typically $\sim \negthickspace200$~$\mu$V), the
QW enters the steady regime, where the temporal variation and the hysteresis in
$G$ disappear.

Now we discuss the excess noise $S$ as a function of $V_{sd}$, which were
measured simultaneously with $G$ as shown in Fig.~2(b).  $S$ is obtained by
subtracting the thermal noise from the total current noise spectral
density~\cite{HashisakaRSI2009}. Around $V_{sd} = 0$~V, the QW shows no excess
noise because of the dissipation-less QH edge transport.  In contrast, finite
excess noise is generated in the nonequilibrium regimes. In the transient
regime, $S$ as well as $G$ show temporal variation.  Remarkably, when the QW
enters into the steady nonequilibrium regime at $|V_{sd}| \gtrsim 200~\mu$V, the
excess noise linearly increases as $\vert V_{sd} \vert$ increases.

The shot noise is characterized by the Fano factor defined by $F \equiv S /
2eI$, where $e$ is the electron charge and $I$ is the current.  In Fig.~2(c),
the Fano factors derived from the measured noise $S$ and the measured current
$I$ are shown for $V_g$'s corresponding to the data shown in Figs.~2(a) and
2(b). In the steady regime, the differential conductance is almost constant,
whereas $S$ linearly increases as $\vert V_{sd} \vert$ increases. The obtained
Fano factor is, therefore, almost constant, as shown by horizontal dashed lines
in Fig.~2(c), which is a characteristic signature expected in conventional shot
noise theory~\cite{Blanter2000PR}. In addition, $F$ systematically depends on
$V_g$; as $V_g$ changes from $-1.6$ to $-1.3$~V, namely, as the width of the QW
increases, $F$ monotonically increases from 0.19 to 0.26.

The above experimental result enables us to decompose the spin-dependent
transmission probabilities of the edge channels as
follows~\cite{DiCarloPRL2006,Nakamura2009PRB}.  In the $\nu = 2$ QH states,
there are only two relevant edge channels, namely, spin-up and spin-down
channels, with transmission probabilities of $\tau_{\uparrow}$ and
$\tau_{\downarrow}$, respectively [see Fig.~1(a)].  On the assumption that the
transmission is independent of electron energy, the conductance $G$ and the Fano
factor $F$ are given as $G = G_0 (\tau_{\uparrow} + \tau_{\downarrow} )$ and $F = \left[ \tau_{\uparrow} (1 - \tau_{\uparrow} ) + \tau_{\downarrow} (1 - \tau_{\downarrow} )\right] / (\tau_{\uparrow} + \tau_{\downarrow})$,
respectively (where $G_0 \equiv e^2/h$)~\cite{Blanter2000PR}.  Based on this
model, we deduce $\tau_{\uparrow}$ and $\tau_{\downarrow}$ with the constraint
that $0 \leq \tau_{\downarrow}, \tau_{\uparrow}\leq 1$, and the
Fano-factor-based electron spin polarization $P_F$ defined by $P_F = \vert
\tau_{\uparrow} - \tau_{\downarrow} \vert / (\tau_{\uparrow} +
\tau_{\downarrow}) = \sqrt {2G_0(1-F)/G-1}$ is obtained.  Naturally, in the
(quasi-)equilibrium regime, where no excess noise is present even when
$|eV|>k_BT$, $F=0$ and thus $P_F = 1$.  This is, however, not the case for the
steady regime.  For example, at $V_g = -1.3$~V, we obtained $P_F \sim 0.62$ as
$G = 1.06G_0$ and $F =0.26,$ as shown in Figs.~2(a) and 2(c), respectively.

Figure 2(d) shows $P_F$ as a function of $V_g$.  Remarkably, $P_F$ monotonically
decreases as the width of the QW increases. We confirmed that a similar result
was also obtained for another QW with $2-\mu$m length.  Such a systematic
dependence of $P_F$ on $V_g$ tells us that the electron scattering that causes
the shot noise mainly takes place inside the QW. Furthermore, because the total
spin momentum should be conserved, the reduction of $P_F$ from unity, namely,
finite electron depolarization, suggests the transfer of spin momentum to other
degrees of freedom, at least partially, to the nuclear spin.

Henceforth, we discuss the result of RDNMR to confirm these findings derived
from the shot noise measurement.  We focus on the transient regime, where QW
shows a hysteresis behavior depending on the $V_{sd}$ sweep direction, as shown
in Fig.~3(a)~\cite{WaldPRL1994}.  In accord with this fact, when $V_{sd}$ is
suddenly increased from 0 to $180$~$\mu$V, the conductance gradually increases
from $0.6 e^2/h$ to $0.9 e^2/h$ with a typical time scale of minutes [solid
circles in Fig.~3(b)].  Such behavior was attributed to the effective hyperfine
field of DNP as reported recently~\cite{Corcoles2009PRB}.  Indeed, the observed
temporal variation of $G$ can be almost eliminated by radio frequency (rf)
irradiation with the NMR frequency of $^{75}$As ($36.408$~MHz at $B=5.0$~T)
[open circles in Fig.~3(b)].  The NMR signal was also detected for
$^{69}$Ga. The result provides strong evidence that DNP occurs in the
nonequilibrium regime.

\begin{figure}[tbp]
\center \includegraphics[width=.79\linewidth]{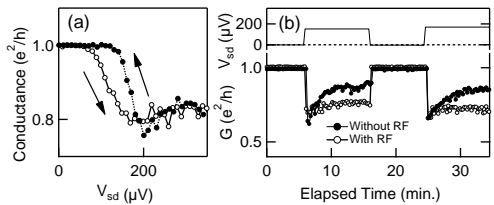} \caption{(a) Typical
differential conductance, where the hysteresis is observed in the transient
regime for $50\lesssim V_{sd} \lesssim 200 \mu$V. (b) Temporal variation of $G$
in the transient regime obtained at $B = 5.0$~T. The data shown by the open
circles are obtained with applying a rf with the NMR frequency for $^{75}$As and
that shown by the solid circles shows the result without rf irradiation.  }
\end{figure}

The Knight shift measurement based on RDNMR is performed to obtain the
Knight-shift-based electron spin polarization $P_K$, which enables us to show
that the above $P_F$ reflects the spin polarization in the QW. The measurement
consists of three steps: ``creation,'' ``irradiation,'' and ``detection''
\cite{Masubuchi2006APL}. First, we set the QW in the transient nonequilibrium
regime to prove that the DNP is created in the QW (creation) via the temporal
variation of $G$.  After waiting a certain time (typically 3 min) for the DNP
development to almost saturate, the QW is set at a given condition and is
irradiated by the rf waves for a few seconds (irradiation).  Finally, the QW is
returned to the initial transient state and the temporal variation of $G$ is
again monitored (detection). By setting the QW at an arbitrary state in the
irradiation step, the NMR spectrum for that state is obtained. If DNP created in
the Creation step is destroyed by the on-resonance rf waves, we detect the
finite temporal variation of $G$ by the redevelopment of DNP in the QW.

The image plot in Fig. 4(a) shows a typical result of the conductance in the
Detection step obtained at $V_g = -1.6$~V and $B=5.0$~T as functions of the RF
frequency and the elapsed time after rf irradiation at $t=0$~s. The QW was set
to the transient regime in the irradiation step. The vertical cross section of
the image plots at $t=10$~s yields the NMR spectrum labeled ``Transient'' in
Fig.~4(b).  We obtained NMR frequencies as the peak value by a Gaussian
fitting. The data obtained for the steady regime are also shown in Fig.~4(b).

\begin{figure}[tbp]
\center \includegraphics[width=.76\linewidth]{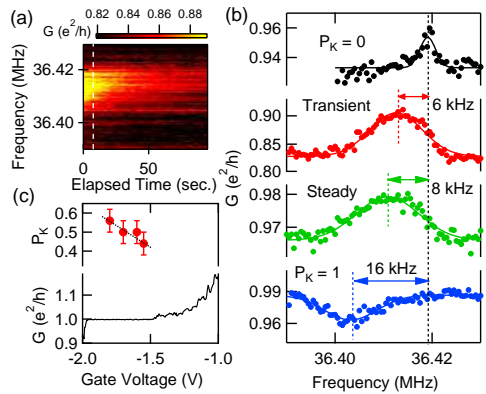} \caption{ (a) Image
plot of the conductance obtained through the RDNMR as functions of the rf
frequency and the elapsed time after rf irradiation.  The QW was set at $V_g = -1.60$~V and at the transient regime at $V_{sd}= 180 \mu$V in the irradiation
step. (b) NMR spectrum obtained at the cross section of (a) at an elapsed time
of 10~s. The solid curve is a Gaussian fit for the spectrum. Also shown is the
NMR spectrum ``Steady,'' in which the QW was set at the steady regime at
$V_{sd}= 280 \mu$V in the irradiation step.  Reference NMR spectra corresponding
to $P_K = 0$ and $P_K = 1$ are shown at the top and the bottom,
respectively. All the data shown in (b) were taken at $V_g = -1.60$~V. (c) $P_K$
obtained by the RDNMR (upper panel) and the equilibrium conductance (lower
panel) as a function of $V_g$. The dotted curve in the upper panel is a guide to
the eye.}
\end{figure}

To derive the Knight shift, the reference frequencies corresponding to $P_{K}=0$
and $P_{K}=1$ are necessary for the same condition.  For $P_{K}=0$, the NMR
spectrum was measured by temporarily depleting the QW in the irradiation step.
The NMR frequency is determined to be 36.419~MHz, as shown in Fig.~4(b), which
corresponds to that of bulk GaAs since there are no conduction electrons that
couple the nuclear spins under rf irradiation.  For $P_{K}=1$, the spectrum is
obtained by setting $V_{sd}=0$ in the irradiation step.  Because only the fully
spin-polarized electrons are present in the equilibrium QW [see Fig.~1(a)], the
NMR frequency gives a reference for the electron polarization of 100\%, which
is determined to be 36.403~MHz, as shown in Fig.~4(b). The signal unexpectedly
shows a dip instead of a peak, which might be relevant to subtle spin dynamics
in the $\nu = 1$ state. Based on the results shown in Fig.~4(b), if we assume
that the Knight shift is proportional to the electron spin polarization, we find
that for the electrons in the QW $P_{K} = 6 \mbox{~kHz}/16 \mbox{~kHz} \sim 0.4$
for the transient regime and $P_{K} = 8 \mbox{~kHz}/16 \mbox{~kHz} \sim 0.5$ for
the steady regime.

The upper panel of Fig.~4(c) shows the polarization $P_K$ obtained in the steady
regime for several $V_g$'s. In the lower panel, the corresponding equilibrium
conductance is shown~\cite{comment1}. Clearly, $P_K$ decreases as the width of
the QW increases (as $|V_g|$ decreases).  This behavior is consistent with that
of $P_F$ obtained via the shot noise shown in Fig.~2(d).  Quantitatively, $P_F$
and $P_K$ are different from each other. This is probably because we simply
assume that the Knight shift is proportional to the electron spin polarization
and neglect that it also depends on the electron density. Nevertheless, the
behavior consistently observed in the two totally different methods supports the
notion that shot noise can serve as a useful tool for detecting electron
polarization as RDNMR does.

The width of the NMR spectra also provides us with useful information on DNP. As
shown in Fig.~4(b), the full width at half maximum (FWHM) for the spectrum of
the depleted QW is $\sim \negthickspace2$~kHz, whereas those of the other
spectra are about 8~kHz. The width that is broader than that of the bulk GaAs is
attributed to the distribution of the electron density in the 2DEG confined in
the heterostructure~\cite{Kuzma1998sci}.  Because no electrons are present in
the depleted QW, no electron distribution exists and the width becomes smaller
than those obtained with electrons present inside the QW.  This observation, as
well as the $V_g$-dependent shot noise, supports the fact that the observed DNP
comes only from the electronic states inside the QW.

Finally, we propose the mechanism of DNP to explain all the experimental
observations as follows.  In the nonequilibrium QW in the QH regime, DNP is
induced by electron scattering between the spin-resolved Landau levels inside
the QW.  In this process, the electron spin flips and nuclear spin flips through
the hyperfine interaction. The excited electrons tunnel in the localized second
Landau levels to escape the QW, providing a partition process of the conduction
electrons at the QW to generate the shot noise.  The unoccupied second Landau
levels are only present inside the QW.  Therefore, as the QW becomes wider, the
number of unoccupied Landau levels available for electron scattering increases,
leading to the large depolarization shown in Figs.~2(d) and 4(c).

In conclusion, we observed shot noise induced by the electron-nuclear
interaction and the resultant electron depolarization in a nonequilibrium
QW. Although so far the RDNMR method has been mainly used to electrically
investigate the DNP in a semiconductor, the present work proves that shot noise
is also useful since it directly probes electron-nuclear spin-flip scattering.
Further investigations in this direction will open a different way to electrically manipulate the coherence of local nuclear spins in semiconductor nanostructures.

We appreciate fruitful discussions with M. Kawamura, N. Kumada, T. Nakajima,
Y. Tokura, and K. Muraki. This work is partially supported by the JSPS Funding
Program for Next Generation World-Leading Researchers.



\begin{thebibliography}{99}
\bibitem{comment2} We confirmed that similar experiments for
other two devices (2-$\mu$m-long QW and quantum point contact) give the result
consistent with that discussed here.
\bibitem{comment1} The RDNMR was performed for the same QW, the results of which
are shown in Figs.~1 and 2, but in a different cooldown. Because of this, the
$V_g$ dependence of $G$ is not the same between Fig.~1(b) and Fig.~4(c). We have
checked that the noise measurement gives a consistent result in this cooldown
for RDNMR.
\end{thebibliography}
\end{document}